\documentclass[a4paper]{article}

\usepackage{INTERSPEECH2020}

\usepackage{times}
\usepackage{soul}
\usepackage{url}
\usepackage[hidelinks]{hyperref}
\usepackage[utf8]{inputenc}
\usepackage{graphicx}
\usepackage{amsmath}
\usepackage{amsthm}
\usepackage{booktabs}
\usepackage{algorithm}
\usepackage{algorithmic}
\usepackage{subcaption}
\title{X-TaSNet: Robust and Accurate Time-Domain Speaker Extraction Network}
\name{Zining Zhang$^{1,2}$, Bingsheng He$^2$, Zhenjie Zhang$^1$}
\address{
  $^1$Singapore R\&D, Yitu Technology\\
  $^2$School of Computing, National University of Singapore}
\email{zining.zhang@yitu-inc.com, hebs@comp.nus.edu.sg, zhenjie.zhang@yitu-inc.com}

\begin{document}

\maketitle
\begin{abstract}
Extracting the speech of a target speaker from mixed audios, based on a reference speech from the target speaker, is a challenging yet powerful technology in speech processing. Recent studies of speaker-independent speech separation, such as TasNet, have shown promising results by applying deep neural networks over the time-domain waveform. Such separation neural network does not directly generate reliable and accurate output when target speakers are specified, because of the necessary prior on the number of speakers and the lack of robustness when dealing with audios with absent speakers. In this paper, we break these limitations by introducing a new speaker-aware speech masking method, called X-TaSNet. Our proposal adopts new strategies, including a distortion-based loss and corresponding alternating training scheme, to better address the robustness issue. X-TaSNet significantly enhances the extracted speech quality, doubling SDRi and SI-SNRi of the output speech audio over state-of-the-art voice filtering approach. X-TaSNet also improves the reliability of the results by improving the accuracy of speaker identity in the output audio to 95.4\%, such that it returns silent audios in most cases when the target speaker is absent. These results demonstrate X-TaSNet moves one solid step towards more practical applications of speaker extraction technology.
\end{abstract}

\section{Introduction}\label{sec:intro}

The separation of speech audios containing different speakers is recognized as one of the core problems in speech processing in the last few decades \cite{cherry1953some,brown1994computational}. Speaker extraction \cite{wang2018voicefilter} is a special case of speech separation, in which the system is expected to regenerate the speech of one particular target speaker from the input audio. Specifically, given a reference audio from the target speaker and a mixed audio with different speakers, the algorithm extracts vocal features from the reference audio, and outputs a new audio clip based on the mixed audio, containing speech from the target speaker only.

Traditional approaches \cite{hershey2016deep,chen2017deep} of speech separation mostly target the spectrogram of the mixed audio based on Short-Term Fourier Transform (STFT). The key is to build a mask over the 2-dimensional spectrogram image, such that irrelevant information to the target speaker is filtered. The magnitude information in spectrograms is incomplete for speech separation tasks, when the phase information of the signals is totally discarded during STFT. The performance of mask-based approaches over spectrogram is known to be bounded by the performance of \emph{optimal} masks based on ground truth, such as Ideal Binary Mask (IBM) \cite{wang2005ideal}, Ideal Ratio Mask (IRM) \cite{li2009optimality}, and Winener Filter-like Mask (WFM) \cite{erdogan2015phase}. It is therefore straightforward to apply these strategies over the raw time-domain waveform instead of the time-frequency domain spectrogram. TaSNet \cite{luo2018tasnet,luo2019conv}, for example, is one of the most successful neural networks for time-domain speech separation, which generates masks over original signals to split waveform based on a given number of speakers. Despite the huge success of TaSNet on speaker separation task, it is unfortunately challenging to directly extend TaSNet for speaker extraction task, due to the following limitations.

\begin{figure*}[t]
    \centering
    \includegraphics[width=6in]{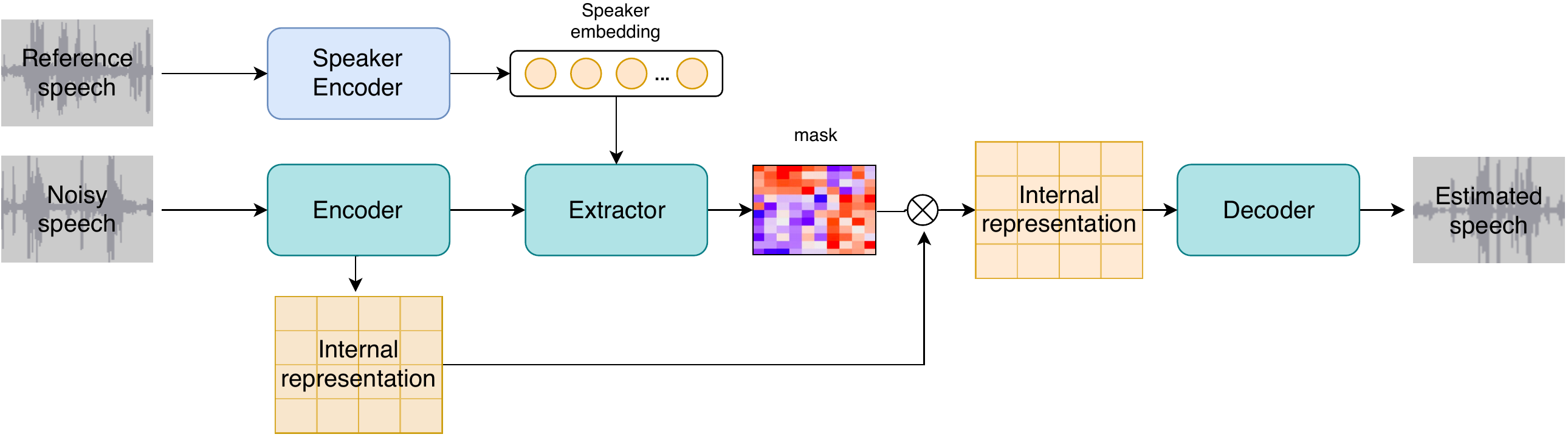}
    \vspace{-5pt}
    \caption{The overall workflow of X-TaSNet.}
    \label{fig:workflow}
    \vspace{-10pt}
\end{figure*}

Firstly, TaSNet does not attempt to identify the speakers during the separation process, and is consequently unable to focus on a specified speaker.
Secondly, TaSNet unrealistically assumes that the model knows the number of speakers as prior knowledge.

In order to maximize the power of such a time-domain approach for the speaker extraction task, one simple solution to the \emph{speaker identification} problem above is to add an extra speaker verification module over the outputs of the speech separation network. It is expected to recognize the output speech from the target speaker. This strategy, however, has difficulties when the target speaker is not present in the mixed audio. Moreover, TaSNet still needs to know the \emph{number of speakers} in advance, which is almost impossible in most of the real-world application scenarios.

In this paper, we propose X-TaSNet. It aims to transfer the knowledge of a speaker verification model to the speech separation model, such that separation and speaker verification are accomplished with a single shot. 
It obtains the speaker identification capability from a pre-trained speaker verification model, and finally includes it in separation network in order to perform speech extraction without knowing the exact number of speakers, even on noisy audio clips.
As a short summary, the core contributions of the work include:

\begin{enumerate}
\item
We present the first time-domain speaker extraction approach, which is seamlessly combined with a speaker verification model. The model exhibits state-of-the-art performance.
\item 
We incorporate novel loss function and corresponding alternating training scheme to fully exploit the power of the time-domain neural network.
\item
We propose new metrics to better measure the accuracy of extracting the voices from the correct target speaker.
\item
We explore a new scenario where the target speaker is not present in the input speech, and propose a new training scheme for this scenario and new metrics for measuring the performance in this case.
\end{enumerate}

The rest of the paper is organized as follows. Section \ref{sec:model} introduces the workflow and the neural network model used in our speaker extraction method. Section \ref{sec:train} discusses our new training scheme to fully exploit the power of the model. Section \ref{sec:experiment} presents the empirical results of our approach over different speech datasets. Section \ref{sec:related} reviews the existing studies and finally Section \ref{sec:conclusion} concludes the paper and discusses the future research directions.

\section{Model}\label{sec:model}

Given a single channel audio clip, denoted by a sequence $x(t)$ over time index $t$, and a reference speech audio $r_i(t)$ from a known speaker $i$, the goal of speech extraction is to generate a new audio clip $s_i(t)$ from $x(t)$, such that $s_i(t)$ contains pure speech audio from speaker $i$.

In this section, we present the details of our X-TaSNet model, which is built on top of the convolutional implementation of TaSNet, i.e., Conv-TaSNet \cite{luo2019conv}. Different from original Conv-TaSNet, our X-TaSNet revises its original encoder, extractor, and decoder, and adds an additional \textit{Speaker Encoder} to the model.
Following \cite{luo2019conv}, the input mixture audio signal is firstly split into overlapping segments of dimension $L$.
These segments are then transformed to vectors of dimension $N$. This transformation $\mathbb{R}^L\xrightarrow{}\mathbb{R}^N$ is learnable, and the mask prediction is performed in $\mathbb{R}^N$ space instead.
After the model completes the transformation, the outputs from \textit{Encoder} are concatenated with the voiceprint of the target speaker on each segment.
The voiceprint is produced based on the reference speech audio $r(t)$ from the target speaker's by using a pre-trained speaker verification model. It is usually called the \textit{Speaker Encoder}. The concatenation is then fed to the \textit{Extractor} to infer the mask for the target speaker's speech.
By applying the mask over the output of the \textit{Encoder}, the masked vectors are the internal representation of the target speaker's speech within each segment in the space $\mathbb{R}^N$. The \textit{Decoder} is then invoked to convert it to the space of $\mathbb{R}^L$ and finally produce the output waveform.

The \textit{Extractor} employs Temporal Convolutional Network (TCN), following \cite{luo2019conv}.
Speaker embedding from \textit{Speaker Encoder} is generated by using Generalized End-to-End (GE2E) speaker verification model \cite{wan2018generalized}.
We skip the details of TCN and GE2E due to space limitation.

\section{Training}\label{sec:train}
To fully unleash the power of the proposed neural network structure in X-TaSNet, we design a few novel technologies on the training scheme based on the model structure.

\noindent\textbf{Additional Loss on Distortion}
While speech separation models, e.g., TasNet, are clearly designed to minimize the loss over all output speakers when the number of speakers is known, it becomes tricky when the speech extraction model targets one particular speaker only. The present speakers in the audio, except the target speaker, are called \emph{distortion speakers} in the rest of the section.
Instead of optimizing the quality of speech from the target speaker only, we find it is equally helpful to minimize the error on distortion speakers.
The core challenge here is how to define the loss function when the number of distortion speakers is unknown to the speech extraction model.
Our solution in X-TasNet is to adopt a new Loss on Distortion (LoD) configuration. Based on the configuration, X-TaSNet generates two outputs. The first output is expected to contain the target speaker's voice only, while the second output is the mixture of all the distortion speakers' voices. 
The LoD is defined as the reconstruction error over the mixture of speech audio from all distortion speakers. This strategy encourages the model to generate a clean separation more than a pure extraction.

\begin{table}[h!]
    \centering
    \caption{Performance of SI-SNRi and NSR with and without LoD. SI-SNRi stands for scale-invariant signal-to-noise ratio improvement, and NSR stands for the Negaitve SI-SNRi Rate. Both of the metrics are explained in Section~\ref{sec:experiment}.}
    \label{tab:LoDmetric}
    \begin{tabular}{l|c|c}
        \hline
          & SI-SNRi &  NSR \\
        \hline
        X-TaSNet w.o. LoD & 12.7  & 6.3\% \\
        X-TasNet w. LoD   & 12.8  & 7.0\% \\
        \hline
    \end{tabular}
\end{table}

In Table~\ref{tab:LoDmetric}, we report the effectiveness of LoD strategy on output speech quality in metrics of SI-SNRi and negative SI-SNRi rate (NSR). SI-SNRi is the scale-invariant signal-to-noise improvement, and NSR is the rate of SNRi that is negative, indicating the likelihood of extremely poor performance usually caused by wrong speaker identity in the output audio. The results imply that LoD does not enhance the average quality of the output speech, and the reason is the downgraded capability on speaker identification.

We believe the additional term of loss on the distortion speakers, which adopts the speech separation losses, is helpful for improving the purity of speech output from the extraction. However, this additional loss may also \emph{distract} the optimization of the model, and confuse the model on extracting the correct speaker. More discussions and empirical evaluations on the robustness issue are available in Section~\ref{sec:experiment}.

\noindent\textbf{Alternating training} According to the observations in Table \ref{tab:LoDmetric}, direct optimization over SI-SNRi may not be the right choice for model training, if we seek to build a robust speech extraction model.
This motivates us to deploy a different training scheme, which targets at improving the accuracy of speaker extraction, namely rates of extracting the correct speakers. This leads to the design of an \emph{alternating training scheme} to replace standard training in the original design of TasNet.

Each training tuple in the dataset is formulated as $\langle x(t),\; r_i(t),\; s_i(t),\; m_i(t) \rangle$ where $x(t)$ is the input audio, $s_i(t)$ is the ground truth of pure speech from target speaker $i$, $m_i(t)$ is the ground truth of the mixture of distortion speakers except speaker $i$, and $r_i(t)$ is the reference speech audio of the target speaker $i$. 

In our new alternating training scheme, the extracting target of one tuple is expanded to all the speakers in the mixed speech.
To be precise, the expanded  training tuple is reformulated as $\langle x(t), \left(r_1(t),s_1(t), m_1(t)\right),\ldots,\left(r_n(t),s_n(t),m_n(t)\right)\rangle$ where $n$ is the total number of speakers in the mixed audio clip $x(t)$.
For each training tuple, the model expands to $n$ voices based on references speech audio clips in $\left(r_1(t),\ldots,r_n(t)\right)$ for all speakers. The corresponding target voices are $\left(s_1(t), \ldots, s_{n}(t)\right)$.
The distortion voice is calculated by summing up all speech signals from other speakers.

Alternating training may look similar to traditional data augmentation strategy commonly used in machine learning training. We believe it unveils a deep insight into the trade-off on training efficiency:
Whether we take more time on exploring different mixed speeches (without alternating training) or to take $n$ steps for each mixture to achieve higher accuracy of pairing up each speaker embedding and corresponding audios (with alternating training).
Given that speaker matching accuracy is the bottleneck under LoD, alternating training is expected to help it to achieve higher performance by increasing the accuracy of extracting the correct speaker's voice.

\noindent\textbf{Speaker Presence Invariant Training (SPIT)}
During the construction of the training dataset, for each mixed audio, we ensure there is at least one reference audio from a speaker not present in the mixed audio $x(t)$. This helps to force the model to consider cases when the target speaker is not present in the audio.
This additional reference audio is denoted as $\hat{r}$.
The target speech for this absent speaker is a segment of silent audio. In order to control the influence of this silent training target, only a small portion of the training tuples include such an absent target speaker.
A special training loss is designed in Section~\ref{subsec:trainloss} in order to penalize the outputs when they are far from silence.

\noindent\textbf{Training loss}\label{subsec:trainloss}
Regarding the training loss, the Scale-Invariant Signal-to-Noise ratio (SI-SNR) is used in the proposed model. It is similar to the Signal-to-Distortion Ratio (SDR) which employed by the original version of TaSNet. It is formulated as:
\begin{equation}\label{eq:1}
SI\text{-}SNR:=
10\log_{10}
\frac
{
    \left\lVert
    \frac
    {
        \langle \hat{s},s \rangle s
    }
    {
        \left\lVert s \right\rVert^2
    }
    \right\rVert^2
}
{
    \left\lVert
    \hat{s}-
    \frac
    {
        \langle \hat{s},s \rangle s
    }
    {
        \left\lVert s \right\rVert^2
    }
    \right\rVert^2
}
\end{equation}
where $s,\hat{s} \in \mathbb{R}^T$ are target speaker's voice, and estimated voice normalized at zero mean. $T$ is the audio length.
The loss function is thus defined as:

\begin{equation}\label{eq:2}
 L:=-SI\text{-}SNR
\end{equation}

When \emph{Loss on Distortion} is activated, the loss function is revised as follows:

\begin{equation}\label{eq:3}
L':=-(SI\text{-}SNR_{target}+SI\text{-}SNR_{distortion})
\end{equation}

If alternating training is employed, the generation of the training dataset follows the strategy introduced above, each random combination of speech audios renders $n$ different target speech audio, based on the number of engaged speakers in the mixed audio.

Finally, by using \textit{Speaker Presence Invariant Training(SPIT)}, there is one more speaker $\hat{s}$ add into the target speaker speech for each mixed audio, such as $\hat{s}$ is not in any of the $n$ present speakers.
Note that the target speech audio of the absent speaker is a silent audio.
Applying the loss as in Equation~\ref{eq:1} directly to silent audios is not feasible, since the norm of them is zero.
To solve this problem, the loss function is revised when the target speech audio is silent, as:
\begin{equation}\label{eq:5}
Decibel:=
10\log_{10}
(\left\lVert \hat{s} \right\rVert^2)
\end{equation}

\section{Experiments}\label{sec:experiment}

\noindent\textbf{Experimental Setup} We train the speaker encoder model GE2E over LibriSpeech \cite{panayotov2015librispeech}, VoxCeleb1 \cite{nagrani2017voxceleb}, and VoxCeleb2 \cite{chung2018voxceleb2} datasets.
For the speech extraction model, the training dataset is taken from LibriSpeech. The mixture is produced by randomly mixing speeches from two different speakers, following the setting in the experiments of Voicefilter \cite{wang2018voicefilter}. we use the same mixture dataset from Google's release\footnote{\url{https://github.com/google/speaker-id/tree/master/publications/VoiceFilter/dataset/LibriSpeech}}. Due to the limited computation resources, we only use the \emph{clean} subset of LibriSpeech.
The audios are clipped to 3 seconds each for more efficient training.
All the reference audios do not appear in mixed audios.
For GE2E and TCN, we follow the settings of the models and training as proposed in \cite{wan2018generalized} and \cite{luo2019conv}.

\noindent\textbf{Evaluation Metrics} Scale-Invariant Signal-to-Noise ratio improvement(SI-SNRi) and Signal-to-Distortion Ratio improvement(SDRi) are used to measure the quality of the extracted speech.

Some models may render higher SDRi over another model, but achieves lower accuracy on extracting the correct target speaker. To better address the robustness of extraction output, we measure speech extraction accuracy using two metrics, including \textit{Negative SI-SNRi Rate} (NSR) as an objective metric and \textit{Speaker Error Rate} (SpkER) as a subjective metric. Specifically, NSR is the portion of the extracted speeches that have negative SI-SNRi, which means that, the extraction doesn't provide quality improvements on the given noisy speeches. Our subjective observations imply that once the model extracts a wrong speaker's voice, SI-SNRi is almost a significant negative value. It is thus a good approximation to the speaker extraction error rate.
Finally, we also report SpkER, which is the error rate of speakers directly evaluated by humans, by listening to the extracted speech audio from the algorithms. 

\noindent\textbf{Quality Evaluation}
In Table~\ref{tab:metrics}, LoD stands for \textit{Loss on Distortion}, AT stands for \textit{Alternating Training}. Voicefilter is adopted as our baseline approach in this group of experiments\footnote{\url{https://github.com/mindslab-ai/voicefilter}}.
Both X-TaSNet and VoiceFilter are trained with the same training and testing data.
Our results show that X-TaSNet outperforms VocieFilter by a large margin, when both are tested using the same speaker verification model.\footnote{Demos can be found at \url{https://speech-ai.github.io/xtasnet/}}
The SDRi and SI-SNRi of X-TaSNet are 2 times better than those of Voicefilter. LoD and alternating training together give the best performance. 
Alternating training, as is shown in Table \ref{tab:metrics}, helps LoD to improve the SDRi and SI-SNRi, as well as the speaker extraction accuracy.
Alternating training without LoD does not generate equally good results, because of the poor data efficiency as discussed in Section \ref{sec:train}.
Table~\ref{tab:metrics} also presents similar NSR and SpkER scores, where SpkER is the subjective metric that measures the speaker extraction error rate. 
This verifies that NSR is a reasonable indicator of speaker extraction error without human efforts on annotations.

\begin{table}[t]
    \centering
    \caption{The performance of VoiceFilter and our proposaed models on mean of SI-SNRi (dB), mean of SDRi (dB), NSR and SpkER.}
    \label{tab:metrics}
    \begin{tabular}{p{3.5cm}|p{.7cm}|p{.7cm}|p{0.6cm}|p{0.6cm}}
        \hline
        Model & SDRi & SI-SNRi & NSR & SpkER \\
        \hline
        VoiceFilter & 7.4 & 6.4 & 9.2\% & 9.5\% \\
        \hline
        X-TaSNet w.o. LoD AT & 13.8 & 12.7 & 6.3\% & 6.6\% \\
        X-TaSNet w.o. AT & 13.8 & 12.8 & 7.0\% & 7.0\% \\
        X-TaSNet w.o. LoD & 14.0 & 13.1 & 5.2\% & 5.5\% \\
        X-TaSNet & \textbf{14.7} & \textbf{13.8} & \textbf{4.3\%} & \textbf{4.6\%} \\
        \hline
    \end{tabular}
\end{table}

\begin{figure}[t]
    \centering
        \includegraphics[width=.94\linewidth]{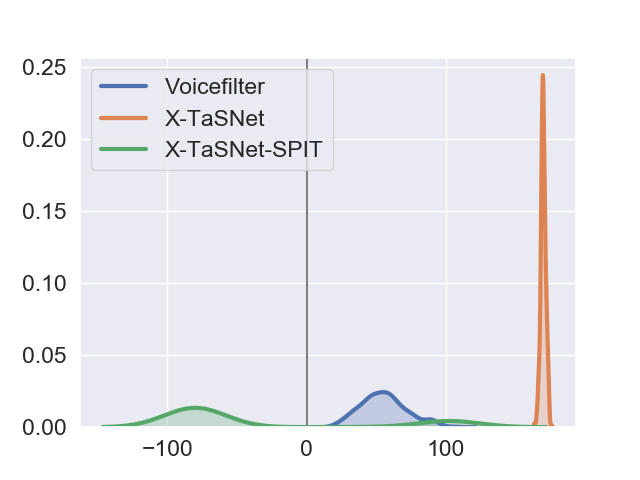}
        \vspace{-5pt}
        \caption{The distribution of extracted absent speakers' voice energy in dB. Voicefilter and X-TaSNet have scale difference since our measure is not scale-invariant. X-TaSNet returns silent audio over most of the arbitrary audios.}
        \label{subfig:dist-db}
        \vspace{-10pt}
\end{figure}

\noindent\textbf{Effects of Speaker Presence Invariant Training}
To better understand the robustness of the models on the extraction of the absent target speaker, we further investigate the energy measurement in dB as defined in Equation~\ref{eq:5} over outputs of the models on arbitrary audio input without the presence of the target speaker. The distribution of the energy of the models is plotted in Figure~\ref{subfig:dist-db}. It is clear that X-TaSNet using Speaker Presence Invariant Training(SPIT) is able to detect the target speaker's absence, and output silent audios with the energy around -100dB.
To have a quantitative evaluation of the speaker's presence dependence property, we propose a new metric Negative Energy Rate (NER), which indicates speaker absence detection accuracy. The results in Figure~\ref{subfig:dist-db} imply that \emph{zero point} could be a good separation boundary for analysis of absence detection accuracy.

With SPIT, X-TaSNet achieves higher NER. But SI-SNRi, which is \textit{12.9}, is below the current best model without SPIT. This is partially because of the higher speaker extraction error which is 8.3\%. Another important reason is about the SI-SNRi distribution.
For X-TaSNet, once it extracts the wrong speaker, there is still some voice in the extracted speech, but for X-TaSNet-SPIT, it may mistakenly decide that the target speaker is not present, and output silent audios.
Silent audios make SI-SNRi a negative number with large absolute value. Thus in the speaker absence scenario, SI-SNRi is not a fair metric.

From the discussion above, we propose a new metric Silence-Invariant Scale-Invariant Signal-to-Noise Ratio improvement(SISI-SNRi). When SI-SNRi turns negative, it is highly likely that the extraction output is on the wrong speaker.
Because we are not interested in the output speech audio quality when the target speaker is wrong in the first place, we combine SISI-SNRi and NSR and build a new metric, which is expected to better reflect the actual usefulness of the speech extraction in real-world application scenarios.
Specifically, the new metric SISI-SNRi denotes the cleanness of the output speech at SI-SNRi when the model targets the correct speaker. 
The results of SISI-SNRi are summarized in Table~\ref{tab:allmetric}.
When compared against X-TaSNet, X-TaSNet-SPIT achieves comparable SISI-SNRi but worse NSR. Although the speaker extraction accuracy degrades, X-TaSNet-SPIT gains the ability of target speaker absence detection. Both Voicefilter and X-TaSNet has 0\% NER, while X-TaSNet-PIT hits 72.4\% accuracy on absence detection.

\begin{table}[h!]
    \centering
    \caption{Performance comparison between Voicefilter, X-TaSNet and X-TaSNet-SPIT}
    \label{tab:allmetric}
    \begin{tabular}{c|c|c|c}
        \hline
        Model & SISI-SNRi & NSR & NER \\
        \hline
        Voicefilter & 7.31 & 9.2\% & 0\% \\
        X-TaSNet & \textbf{15.57} & \textbf{4.3\%} & 0\% \\
        X-TaSNet-PIT & 14.50 & 8.3\% & \textbf{72.4\%} \\
        \hline
    \end{tabular}
\end{table}

\section{Related Work}\label{sec:related}

Speech extraction is a task closely related to speech separation.
Due to the limitations of order unawareness in speech separation, as well as the progressive improvements of neural speaker encoder, researchers are attempting to specify the speaker embedding in order to target the speech audio from specific speakers.

\cite{vzmolikova2017learning,zmolikova2017speaker} are models extracting the target speaker's voice from an array of microphones. They use mask-based methods, and train the speaker information encoder jointly with the model.
\cite{delcroix2018single} uses a similar method, and proves that the method is feasible for the single-channel scenario.
\cite{wang2018deep} also solves on single-channel speaker extraction, but with short reference utterances. It achieves this by creating embedding of mixture and reference utterance separately and combined to a single input to mask prediction network. 
\cite{zmolikova2019speakerbeam} proposed SpeakerBeam, which discusses different ways of utilizing the reference audio information, 
and \cite{delcroix2019compact} is a simpler version of SpeakerBeam with fewer parameters but comparable performance.
\cite{xiao2019single} uses an attention network, which is different from the above-mentioned models. However, it is only used in the scenario where the inventory of speakers is given.
\cite{wang2018voicefilter} is our baseline model. It uses a pretrained speaker verification model to extract the reference speaker's voiceprint. The speaker identification ability is highly dependent on this separate model. The benefit is that, we can use a separate large corpus to train the speaker verification model to increase its generalization ability on speaker identification, without considering the speech extraction model.
All the methods described above process the speech signal on the time-frequency domain, instead of the time domain as X-TaSNet does.

\section{Conclusion and Future Work}\label{sec:conclusion}
In this paper, we present X-TaSNet, a new speech extraction approach based on a time-domain speech separation neural network. By employing new loss functions and training scheme, X-TaSNet demonstrates significant performance enhancement over the state-of-the-art solution, on both output speech audio quality and speaker identity robustness. In the future, we would like to explore two research directions.
Firstly, the NER of X-TaSNet-SPIT is below 80\% which may not be sufficient for serious scenarios. We believe improving on NER is an important direction. Secondly, the current design and testing of X-TaSNet targets at speech separation task, which could be extended to handle other types of noises, such as background music, for other speech enhancement tasks.

\bibliographystyle{IEEEtran}

\bibliography{mybib}

\end{document}